\documentclass[10pt,letterpaper,twocolumn]{article} 

\usepackage{ol2}
\usepackage[draft,implicit=false]{hyperref}
\usepackage{amsmath}

\begin{document}
\newcommand{\comment}[1]{}
\newcommand{\E}{\mathrm{E}}
\newcommand{\Var}{\mathrm{Var}}
\newcommand{\bra}[1]{\langle #1|}
\newcommand{\ket}[1]{|#1\rangle}
\newcommand{\braket}[2]{\langle #1|#2 \rangle}
\newcommand{\be}{\begin{equation}}
\newcommand{\ee}{\end{equation}}
\newcommand{\ba}{\begin{eqnarray}}
\newcommand{\ea}{\end{eqnarray}}
\newcommand{\SD}[1]{{\color{magenta}#1}}
\newcommand{\HME}[1]{{\color{green}#1}}
\newcommand{\rem}[1]{{\color{cyan} \sout{#1}}}
\newcommand{\alert}[1]{\textbf{\color{red} \uwave{#1}}}
\newcommand{\Y}[1]{\textcolor{BurntOrange}{#1}}
\newcommand{\R}[1]{\textcolor{red}{#1}}
\newcommand{\B}[1]{\textcolor{blue}{#1}}

\twocolumn[ 

\title{Linear negative dispersion with a gain doublet via optomechanical interactions}


\author{Jiayi Qin,$^{1,*}$ ,Chunnong Zhao$^1$, Yiqiu Ma$^1$, Li Ju$^1$, David G. Blair$^1$}

\address{
$^1$School of Physics, University of Western Australia, WA 6009, Australia
\\
$^*$Corresponding author: jiayiqinphysics@gmail.com
}

\begin{abstract}
Optical cavities containing a negative dispersion medium have been proposed as a means of improving the sensitivity of laser interferometric gravitational wave (GW) detectors through the creation of white light signal recycling cavities. Here we classically demonstrate that negative dispersion can be realized using an optomechanical cavity pumped by a blue detuned doublet. We used an 85mm cavity with an intra-cavity silicon nitride membrane. Tunable negative dispersion is demonstrated, with a phase derivative $d\varphi/df$ from $-0.14$ Deg$\cdot$Hz$^{-1}$ to $-4.2\times10^{-3}$ Deg$\cdot$Hz$^{-1}$.
\end{abstract}


 ] 

{\it Introduction---} In 1839, W.R. Hamilton pointed out that the group velocity of a wave is different from its phase velocity \cite{WRHamilton}. In 1881, Lord Rayleigh \cite{Rayleigh81} commented that a pulse of light travels at the group velocity (rather than the phase velocity) inside a medium. He further developed the theory of anomalous dispersion and first demonstrated anomalous dispersion for mechanical oscillator \cite{Rayleigh2,Rayleigh3}. Later, Sommerfeld and Brillouin theoretically showed that dispersion is anomalous inside an absorption line, where the group velocity exceeds c, the light velocity in the vacuum \cite{Brillouin}.

More recently anomalous dispersion has been investigated in various schemes. In 1970, Garrett and McCumber \cite{Garrett} studied the properties of Gaussian light pulses in absorptive media. In 1971, Casperson and Yariv \cite{Casperson} demonstrated that the velocity of ultrashort pulses through a high-gain medium is a function of the gain, and agrees with the theoretical results. In 1981, Chu et. al \cite{Chu} experimentally verified the theoretical predictions in \cite{Garrett} showing that pulses propagating in samples of GaP:N tuned to the bound A-exciton line propagate with little pulse shape distortion, at a group velocity that either exceeds $c$ or becomes negative. In 1985, Segard and Macke \cite{Segard} demonstrated significant pulse advances $\sim2$ $\mu$s with negligible distortion through a linear molecular absorber at millimeter wave-lengths.

More recently, the negative dispersion was observed in atomic systems via electromagnetical interactions.
The effect is caused by atomic coherence in degenerate two-level systems. It leads to negative dispersion
 at the resonant frequency of an atomic transition. 
  In 1999, Akulshin et. al \cite{Akulshin} realized steep negative dispersion up to $dn/d\nu\simeq-6\times10^{-11}$ Hz$^{-1}$ in coherently prepared Rb Vapor.

In 1994, Steinberg and Chiao \cite{Steinberg} realized that, in a medium with a gain doublet,
 there exists a transparent region with anomalous dispersion between the two gain peaks. In 2000, Wang et. al \cite{LJWANG} reported gain-assisted linear negative dispersion
 in atomic caesium gas.
 In 2011, Safavi-Naeini et. al \cite{SN} reported an optically tunable delay of 50 nanoseconds and superluminal light with 1.4 microsecond signal advance in a nanoscale optomechanical crystal. In 2012, Ivanov et al. \cite{Ivanov} observed the 'fast light' effect in a cryogenic microwave resonator.

The concept of using a negative dispersion medium to improve the broadband sensitivity of gravitational wave detectors was first proposed by Wicht et al \cite{Danzmann} in 1997. In 2007, Salit et al \cite{Shahriar} proposed a practical design for a white-light signal recycling GW interferometer by putting a negative dispersion medium inside the recycling cavity. By taking into account the quantum noise and stability of the system,
Ma et al \cite{Ma} excluded the possibility of using such a double-pumped optomechanical filter for improving the shot-noise-limited sensitivity of GW detectors in the stable regime. In order to achieve sensitivity improvement, one approach is exploring the unstable regime with additional feedback control for stabilization, as proposed by Miao et al \cite{Miao}. The other approach is using multiple control fields, instead of just two, to construct an optical filter that has a different spectral shape, e.g., the one considered by Zhou et al. \cite{Shahriar2} but using atomic systems.

In this paper, we make a proof-of-principle demonstration of using the optomechanical device to realize an active filter which possesses interesting optical property. We report the experimental realization of a gain-assisted linear negative dispersion in an optomechanical system with two blue-detuned control beams. Using an 85-mm optical cavity coupled with a silicon nitride membrane, we
demonstrate optically tunable negative dispersion that is equivalent to replace the cavity with
a negative dispersion medium , which has a phase derivative $d\varphi/df$ from $-0.14$ Deg$\cdot$Hz$^{-1}$ to $-4.2\times10^{-3}$ Deg$\cdot$Hz$^{-1}$. Our result shows the potential possibility of using such optomechanical systems to build white-light signal recycling cavities in laser interferometric gravitational wave detectors for improving the sensitivity subject to the significant reduction of mechanical resonator's thermal noise. Such a free-space optomechanical filter can be an alternative but less lossy compared to atomic systems where the light has to go through the medium \cite{Ma,Miao, Shahriar2, EITloss}. Our demonstration shows the feasibility of using optomechanical interactions to achieve a required optical response.

\begin{figure}[htb]
\centerline{\includegraphics[width=7.5cm]{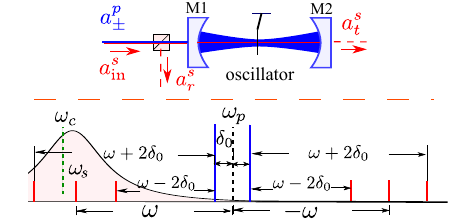}}
\caption{(Color online)Configuration schematics (Top) and frequency relationships. The signal light having frequency $\omega_s$ is injected into an optical cavity with a silicon nitride membrane in the middle which acts as an oscillator at the resonant frequency $\omega_{m}$. The position of the membrane is chosen to introduce a linear optomechanical coupling. The radiation
pressure forces from the beatings between the signal light at the frequency $\omega_{s}=\omega_{p}-\omega$ and the control beams at the frequency $\omega_p\pm\delta_0$ drive the mechanical oscillator which in turns create sidebands destructively interfering with the signal
light, which in effect create linear negative dispersion.}
\label{config}
\end{figure}

{\it Optomechanical dynamics---} Optomechanical interactions can be used to create the linear negative dispersion in the scheme shown in Fig.\ref{config}. Two control beams maintain strong fields $\bar {a}_{\pm}e^{-i(\omega_p\pm\delta_0)t}$ in the cavity at the resonant frequency $\omega_0$. A weak signal light $\delta \hat a_{\rm in}=\hat{a}^s_{\rm in} e^{-i\omega_{s} t}$ is injected to the same port (M1). The frequency differences between the control fields central frequency $\omega_p$ and the signal fields need to be close to the mechanical resonant frequency for effective driving of the mechanical mode.

The Hamiltonian which describes this system is given by:

\be
\hat{H}=\hbar(\omega_{0}+g_{0}\hat{x}) \hat{a}^{\dag}\hat{a}+\hat{H}_{m}+\hat{H}_{\gamma}.
\label{hamiltonian}
\ee
Here, $\hat{H}_{m}=\hat{p}^{2}/2 m+m\omega_{m}^{2}\hat{x}^{2}/2$  is the free Hamiltonian of the mechanical oscillator.
$H_\gamma=-i\hbar\sqrt{2\gamma_1}\hat{a}\hat{a}_{\rm in}^{\dagger}
-i\hbar\sqrt{2\gamma_2}\hat a\hat b_{\rm in}^{\dagger}+h.c$ describes the interaction between the intra-cavity field $\hat{a}$ and external electromagnetic fields $\hat{a}_{\rm in}$ and $\hat b_{\rm in}$ with bandwidths $\gamma_1=cT_1/4L$ and $\gamma_2=cT_2/4L$ through the cavity mirrors M1 and M2 respectively. $g_{0}$ is the linear optomechanical coupling strength.
Since the signal light $\hat a^s_{\rm in}(\omega)$ in our experiment is a classical field, we neglect the vacuum fluctuation term. In the rotating frame at frequency $\omega_{p}$ where $\omega=\omega_p-\omega_s$, we have:

\begin{subequations}\label{eq_2}
\begin{align}
&\hat{x}(\omega_\pm)=-\chi_m^{-1}(\omega_{\pm})\hbar g_{0}[\bar {a}_\mp \hat{a}^{\dag}(\omega\pm2\delta_0)+\bar{a}_\pm \hat{a}^{\dag}(\omega)\nonumber\\
&\qquad\quad+\bar{a}^{*}_{\mp}\hat a(-\omega)+\bar{a}^{*}_{\pm}\hat a(-\omega\mp2\delta_0)],\label{eq eom2}\\
&\hat{a}(\omega)=-i\chi^{-1}_c(\omega)g_{0}[\bar {a}_+\hat{x}(\omega_+)\nonumber\\
&\qquad+\bar{a}_-\hat{x}(\omega_-)]+\sqrt{2\gamma_1}\hat{a}^s_{\rm in}(\omega).\label{eq eom3}
\end{align}
\end{subequations}

where $\omega_{\pm}=\omega\pm\delta_0$ and  $\chi_m(\omega_{\pm})=m(\omega^2_m-\omega_{\pm}^2-i\gamma_m\omega_{\pm})$, $\chi_c(\omega)=[-i(\omega-\Delta)+\gamma]$ are the mechanical and optical susceptibility, respectively. The $\gamma_{m}$ and $\gamma=\gamma_1+\gamma_2$ are the bandwidths of the mechanical oscillator and the cavity, respectively.  We choose the frequency detuning to be $\Delta=\omega_p-\omega_c\sim\omega_{m}$. In our system where $\delta_0\ll\gamma$, We make use of the near resonance approximations: $\omega-\Delta\ll\gamma$, $\omega-\Delta\pm2\delta_0\ll\gamma$. Thus, the optical susceptibilities can be approximately written as: $\chi_c(\omega\pm2\delta_0)\sim\chi_c(\omega)\sim\gamma$.

Since the optical anti-Stokes sidebands $\hat a(-\omega)$ and $\hat a(-\omega\mp2\delta_0)$ are far detuned where $\gamma\ll2\omega_m$, therefore, they can be safely ignored. This approximation is called \emph{resolved sideband approximation}. Here, the Stokes sidebands under the above near resonance approximation can be written as:

\begin{figure}[b]
\centerline{\includegraphics[width=7.5cm]{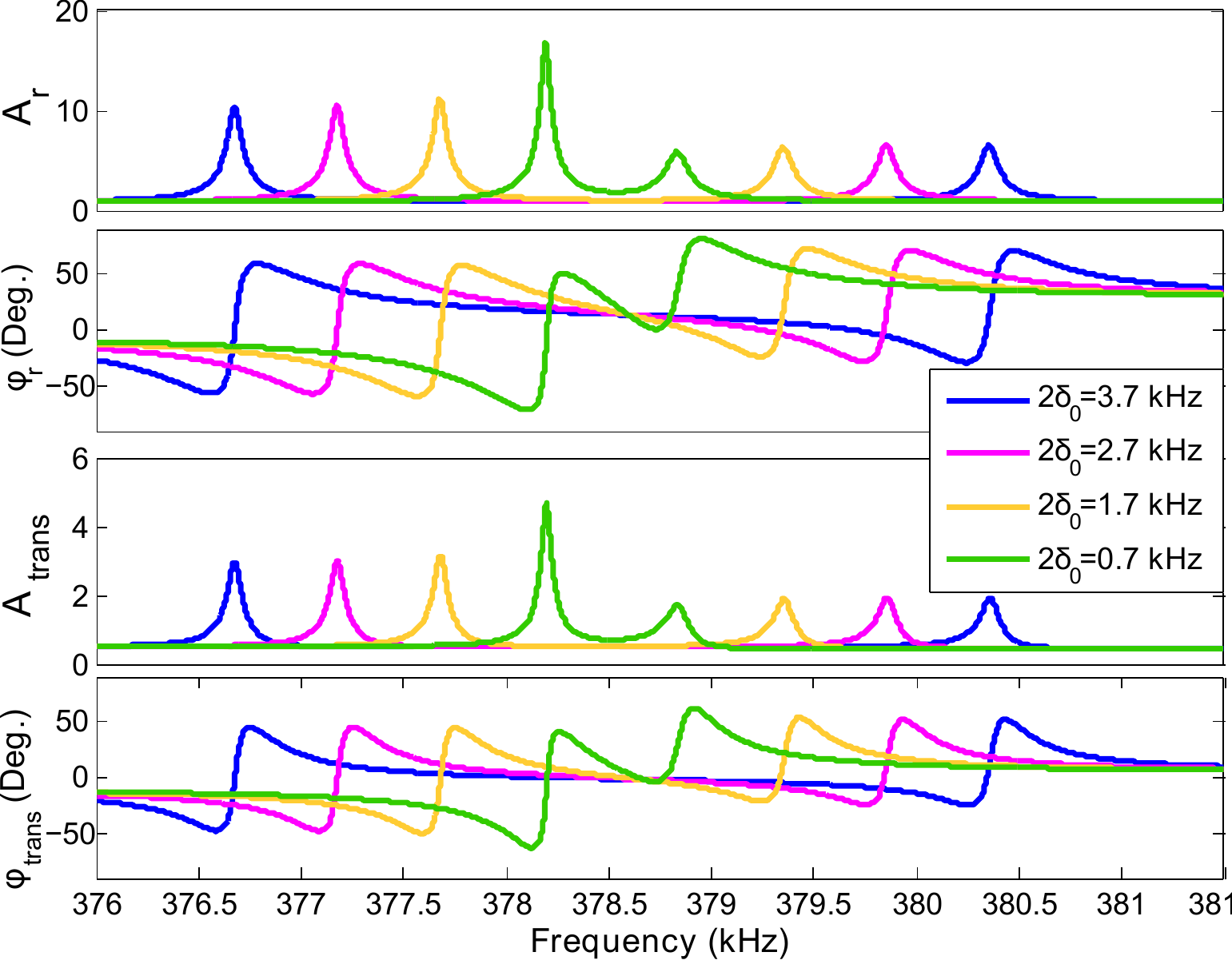}}
\caption{(Color online) Theoretical reflectivity and transmissivity in the double gain scheme: (a) Reflectivity amplitude $A_{\rm ref}$, (b) reflectivity phase $\phi_{\rm ref}$, (c) transmissivity amplitude $A_{\rm trans}$ and (d) transmissivity phase $\phi_{\rm trans}$. The parameters are: $\gamma_m=340 $ Hz, $\Gamma_+=170$ Hz, $\Gamma_-=140$ Hz, $\omega_m=378.55$ kHz, the frequency detuning $\Delta-\omega_m=9$ kHz in $\chi_c(\omega)$.}
\label{fig:theoretical}
\end{figure}

\begin{subequations}\label{eq_3}
\begin{align}
&\hat{a}^\dag(\omega)\approx\frac{i{g}_{0}}{\gamma}[\bar a^*_+\hat{x}(\omega_+)+\bar{a}^*_-\hat{x}(\omega_-)]\nonumber\\
&\qquad\quad+\frac{\sqrt{2\gamma_1}}{\gamma}\hat a^{s\dag}_{\rm in}(\omega),&\label{eq mode0}\\
&\hat{a}^\dag(\omega\pm2\delta_0)\approx\frac{i{g}_{0}}{\gamma}\bar a^*_{\mp}\hat{x}(\omega_{\pm}),\label{eq sidebands}
\end{align}
\end{subequations}

Substituting \eqref{eq_3} into radiation pressure force terms in the equations of motion for mechanical displacement Eq. \eqref{eq eom2} leads to:

\begin{align}\label{eq_4}
&   \begin{bmatrix}
    \chi_{\rm eff}(\omega_+) & i\sqrt{\Gamma_+\Gamma_-}\\
    i\sqrt{\Gamma_+\Gamma_-} & \chi_{\rm eff}(\omega_-)
    \end{bmatrix}
    \times\begin{bmatrix}
    x(\omega_+)\\
    x(\omega_-)
    \end{bmatrix}
   = \begin{bmatrix}
        \beta_+ \\
        \beta_-
    \end{bmatrix}a^{s\dag}_{\rm in}(\omega).
\end{align}

The optomechanical anti-damping rates $\Gamma_{\rm \mp}$ are given by:
\be
\Gamma_{\pm}=\frac{\hbar g_0^2|\bar a_{\pm}|^2}{2m\omega_m\gamma},
\ee
 and can be tuned by changing the control field power. The effective mechanical response function $\chi_{\rm eff}(\omega_\pm)=\chi_{m}(\omega_\pm)/2m\omega_m+i(\Gamma_{+}+\Gamma_-)$ where in the near-resonance case, we have $\chi_m(\omega\pm)\approx-2m\omega_m(\omega_\pm-\omega_m+i\gamma_m/2)$
.  The parameters $\beta_\pm$ are defined as $\beta_\pm\equiv\sqrt{2\gamma_1}\Gamma_\pm/\gamma\bar a^*_\pm$. The beat between the signal field and one control field $\bar a_+$ creates a radiation pressure force that induces the mechanical motion $x(\omega_+)$. This mechanical motion $x(\omega_+)$ then modulates the control fields $\hat a_+$ and $\hat a_-$to produce Stokes sidebands at frequency $\omega$ and $\omega+2\delta_0$, respectively. The anti-damping rates $\Gamma_+$ and $\Gamma_-$ in $\chi_{\rm eff}(\omega_+)$ are caused respectively, by the beat between $\hat a(\omega)$ and the control field $\bar a_+$ and between $\hat a(\omega+2\delta_0)$ and the control field $\bar a_-$. The field at frequency $\omega$ also interacts with the control field $\hat a_-$ to create a radiation pressure force. This force drives the mechanical motion $x(\omega_-)$ and give rise to the interaction terms $\sqrt{\Gamma_+\Gamma_-}$ in \eqref{eq_4}. The terms on the bottom line of \eqref{eq_4} are similarly explained.

\begin{figure}[b]
\centerline{\includegraphics[width=7.5cm]{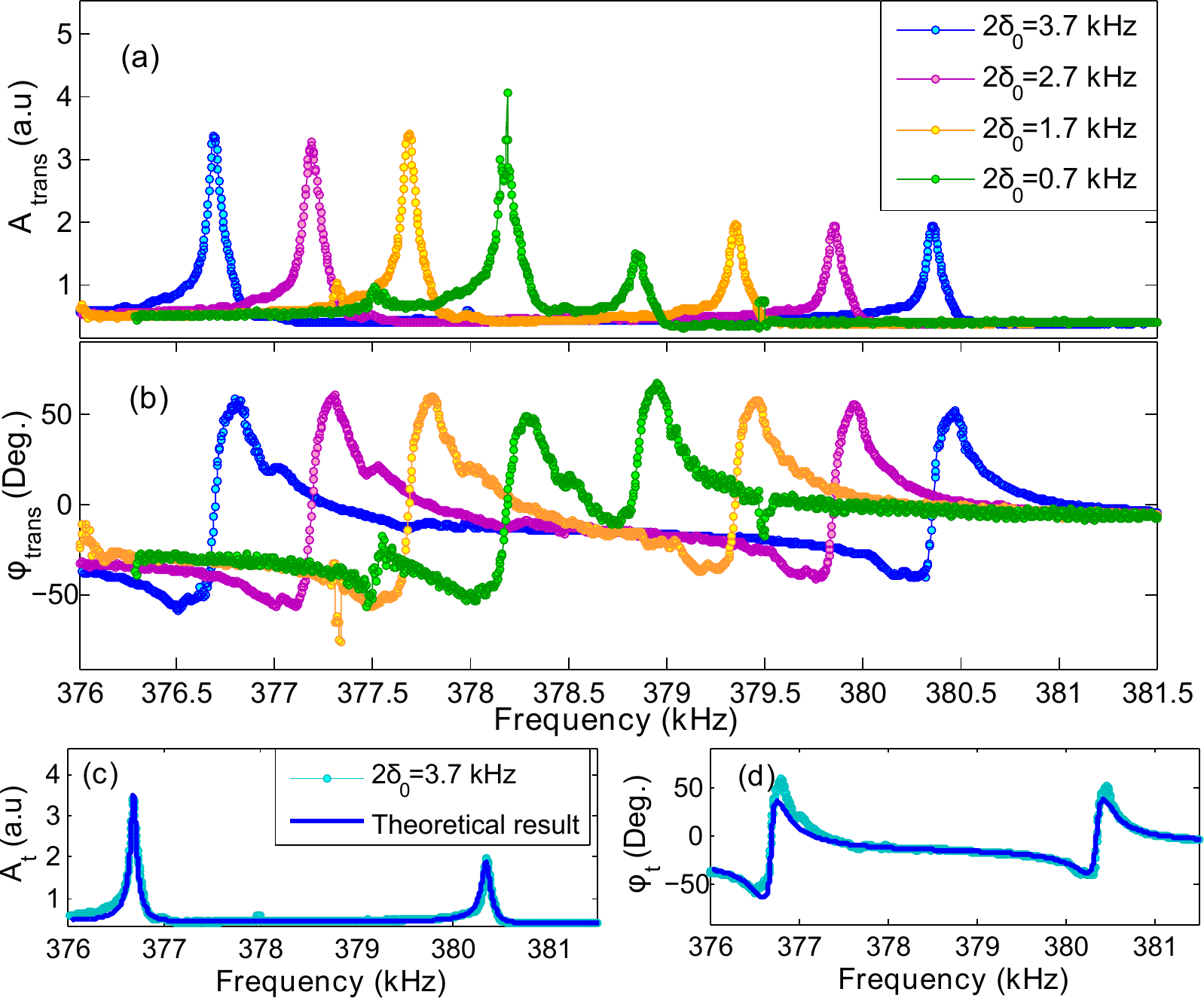}}
\caption{(Color online) Experimental transmissivity in the double gain scheme: (a) $A_{\rm trans}$; (b) $\varphi_{\rm trans}$; (c) and (d) combine transmissivity data and theoretical result at $2\delta_0=3.7$ kHz. The parameters are same as Fig. \ref{fig:theoretical}.}
\label{fig:result}
\end{figure}

Substitute \eqref{eq_4} into \eqref{eq mode0} and use the input-output relations: $\hat a^\dag_{\rm r}=-\hat a^\dag_{\rm in}+\sqrt{2\gamma_1}\hat a^\dag$, $\hat a^\dag_{\rm t}=\sqrt{2\gamma_2}\hat a^\dag$, we have effective reflectivity and transmissivity as:

\begin{subequations}
\begin{align}
&r(\omega)=-1+2\eta_c\frac{1}{(1+i\chi_{\varphi})},&\label{eq r}\\
&t(\omega)=2\sqrt{(1-\eta_c)\eta_c}\frac{1}{(1+i\chi_{\varphi})},\label{eq t}
\end{align}
\end{subequations}

where $\chi_{\varphi}=\Gamma_+/[\omega-\omega_{m}+\delta_0+i\gamma_m-i\Gamma_-]+\Gamma_-/[
\omega-\omega_{m}-\delta_0+i\gamma_m-i\Gamma_+]$ and the cavity coupling parameter is $\eta_c=\gamma_1/(\gamma_1+\gamma_2)$. In Fig. \ref{fig:theoretical}, we give the theoretical reflectivity and transmissivity using parameters which are matched to our measurement results in Fig. \ref{fig:result}. In our experiment, we detect the signals at the transmission port. To build white light signal recycling cavities in gravitational wave interferometers, the reflection output will be used instead of transmission one to reduce quantum noises coupled from the transmission port.

\begin{figure}[htb]
\centerline{\includegraphics[width=7.5cm]{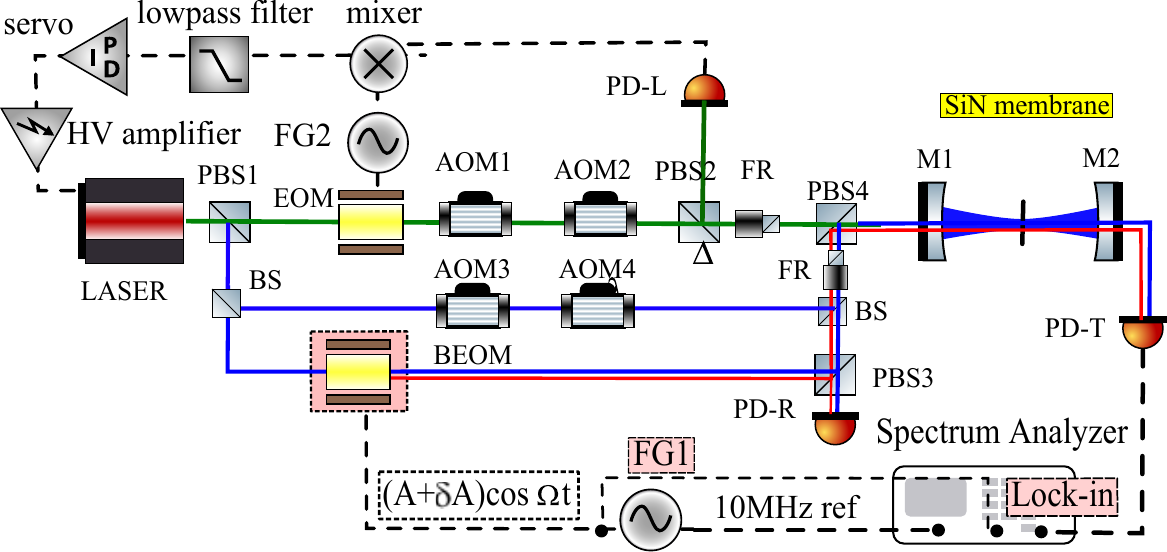}}
\caption{(Color online)Experimental setup. The 85 mm long cavity sits in a vacuum chamber with a central silicon nitride membrane oscillator(1mm$\times$1mm$\times$50nm, effective mass $40$ ng). The green line represents the locking light for stabilizing the laser frequency to the cavity resonance using Pound-Drever-Hall (PDH) locking \cite{pdh}. The blue lines represent the control fields, with  polarization orthogonal to the locking light. The broadband electro-optic modulator (EOM) generates a lower-sideband from the control light, which is our signal light (red line). The frequency differences between the locking and control fields were created using pairs of $80$ MHz AOMs in the optical paths.}
\label{fig:setup}
\end{figure}

{\it Experiment---}
In our experimental setup shown in Fig. \ref{fig:setup}, the key element is an 85 mm high-finesse optical cavity containing a high stress silicon nitride membrane, which has a quality factor of $\sim1.5 \times10^6$ at its fundamental mechanical resonance $\sim378.5$ kHz \cite{omitsup}, measured at vacuum level of $10^{-5}$ mbar. The experiment was conducted at a relatively low vacuum ($\sim10^{-2}$ mbar) to use the gas damping to stabilize the system since the blue detuned control light will create negative-damping. The $Q$-factor was reduced to $\sim10^3$. However, active or passive cooling can be used to stabilize the system and to suppress thermal noise in future practical systems.

Our optical cavity is mounted on an invar spacer machined by electrical discharge machining with accuracy of 0.1 $\mu \rm m$ and fixed in a vibration isolated vacuum tank. The mirrors M1 and M2 are clamped at the ends of the spacer. The membrane frame is bonded onto a piezoelectric actuator with Yacca gum, a natural resin with low intrinsic loss \cite{yacca}. The transmissivity $T_1$ $(T_2)$ of M1 (M2) is $245.1\pm2.8$ ppm ($16.93\pm0.20$ ppm), which was measured in an empty cavity. The experiment was conducted at room temperature using a 1064nm Nd:YAG laser \cite{omit}.

\begin{figure}[htb]
\centerline{\includegraphics[width=7.5cm]{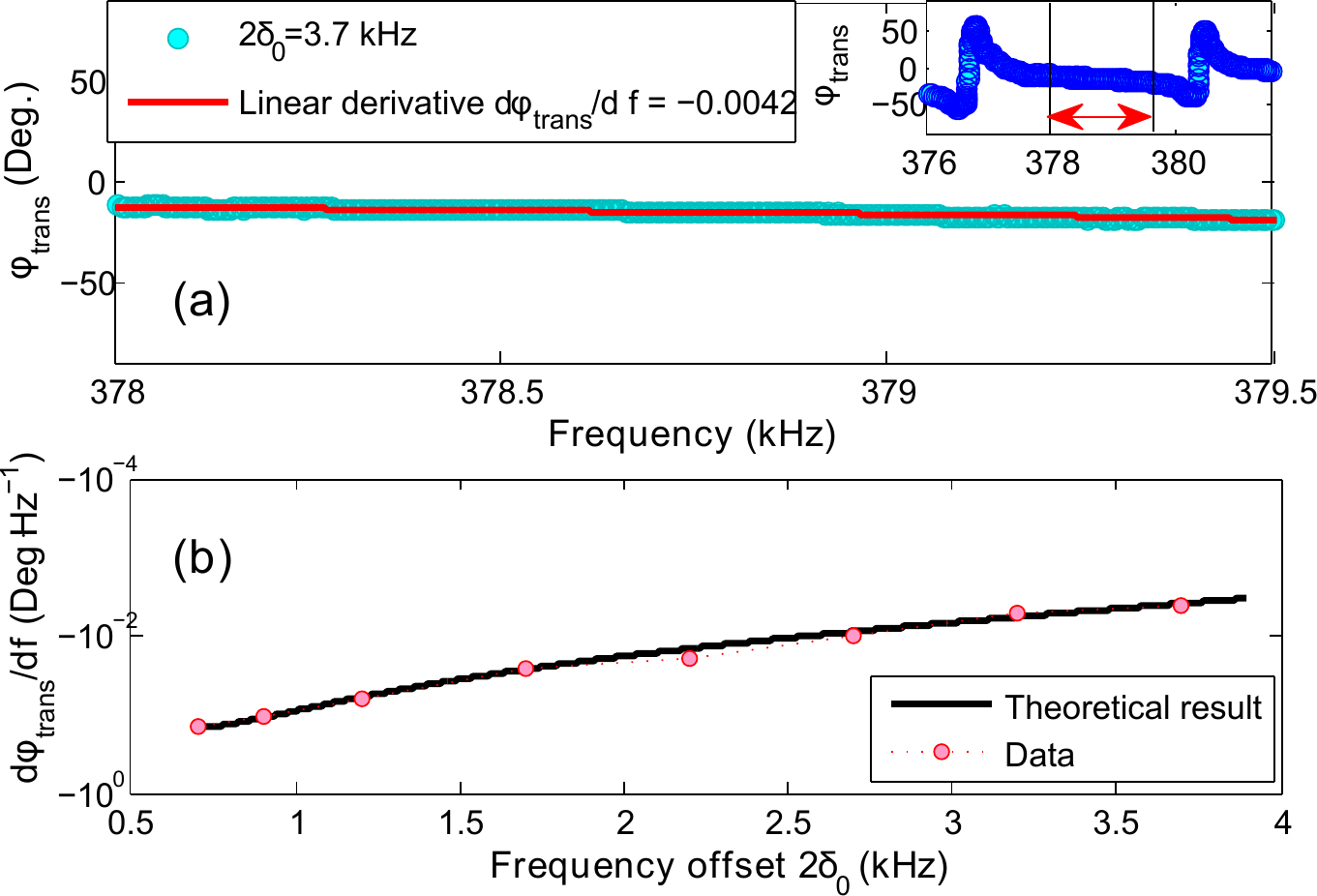}}
\caption{(Color online) (a) Zoomed-in linear transmission phase for frequency offset $\delta_0=3.7$ kHz. (b)Linear derivatives of transmissivity phase v.s. frequency offset $2\delta_0$ at the frequency ($\omega_m-0.28 kHz$).}
\label{fig:phasederivative}
\end{figure}

To create a wide linear negative dispersive region, two blue-detuned control fields at frequencies $\omega_p\pm\delta_0$ were injected into the cavity. The field $\hat a_-$ passes through a broadband EOM to generate the weak signal light $\hat a^s_{\rm in}$. Using a lock-in amplifier\cite{omit}, we measured the transmission of the cavity by detecting the beat signals between the signal field and the control fields at the transmission port. In Fig. \ref{fig:result}, we show experimental results of the transmission amplitude and phase for different frequency offsets $\delta_0$. In Fig. \ref{fig:phasederivative} (a), we show the linear region of transmission phase for frequency offset $\delta_0=3.7$ kHz with linear curve fitting . While there is some small distortions in the phase response at large frequency offsets, the linear negative dispersion region is well behaved within expected error.

In Fig. \ref{fig:phasederivative}(b), we show the first order derivatives of transmission phase at different frequency offsets, which range from $-0.14$ Deg$\cdot$Hz$^{-1}$ to $-4.2\times10^{-3}$ Deg$\cdot$Hz$^{-1}$.

{\it Conclusion---}
 Tunable negative dispersion has been created in an optomechanical cavity with an intra-cavity silicon nitride membrane. Double pumping is used to create a wide linear negative dispersion regime. The system has similar performance to an atomic gas system with a gain doublet and could be a less lossy alternative.
  The technique described here can be extended to broadband sensitivity improvement using double-gain blue-detuned cavities with feedback control \cite{Miao}. Practical systems would require a mechanical resonator with high $Q$-factor to environment temperature ratio, as discussed in \cite{Miao}. Optical dilution can be used to increase mechanical $Q$-factor by a factor of hundred or even more \cite{Ma2}. Our demonstration illustrates the potential of using optomechanical interactions to achieve the required active optical filter response for improving the sensitivity of gravitational wave detectors.

We thank Haixing Miao, Xu Chen, Yanbei Chen and Stefan Danilishin for useful discussions. We thank Gary Light for technical support. We would like to thank the LIGO Scientific Collaboration, The Gingin International Advisory Committee, and our collaborators for useful advice. This research was supported by the Australian Research Council (DP120104676 and DP120100898).

\end{document}